\documentclass[11pt,nofootinbib,floatfix,onecolumn,preprintnumbers]{revtex4}
\pdfoutput=1

\usepackage{amsmath,amssymb}
\usepackage{graphicx}
\usepackage{rotating}
\usepackage{color}
\usepackage{multirow} 
\usepackage{fontenc} 
\usepackage{slashed}
\usepackage{longtable}
\usepackage{hyperref}

\def\hc{\text{h.c.}}

\newcommand{\AddrBonn}{%
Bethe Center for Theoretical Physics and
Physikalisches Institut der Universit\"at Bonn,
Nussallee 12, D-53115 Bonn, Germany }

\newcommand{\AddrIFIC}{%
Instituto de F\'{\i}sica Corpuscular (CSIC-Universitat de Val\`{e}ncia),
Apdo. 22085, E-46071 Valencia, Spain
}

\preprint{IFIC/18-34}
\preprint{BONN-TH-2018-13}
\begin{document}

\title{Lepton Flavor Violation in a $Z^\prime$ model for the $b \to s$ anomalies}

\author{Paulina Rocha-Mor\'an}\email{procha@th.physik.uni-bonn.de}
\affiliation{\AddrIFIC}
\affiliation{\AddrBonn}

\author{Avelino Vicente}\email{avelino.vicente@ific.uv.es}
\affiliation{\AddrIFIC}

\begin{abstract}
In recent years, several observables associated to semileptonic $b \to
s$ processes have been found to depart from their predicted values in
the Standard Model, including a few tantalizing hints of lepton flavor
universality violation.  In this work we consider an existing model
with a massive $Z^\prime$ boson that addresses the anomalies in $b \to
s$ transitions and extend it with a non-trivial embedding of neutrino
masses. We analyze lepton flavor violating effects, induced by the
non-universal interaction associated to the $b \to s$ anomalies and by
the new physics associated to the neutrino mass generation, and
determine the expected ranges for the most relevant observables.
\end{abstract}

\maketitle

\section{Introduction}
\label{sec:intro}

The Standard Model (SM) of particle physics provides a precise
description to a vast amount of phenomena as well as a deep
understanding of the fundamental laws that govern them. However,
despite its outstanding success, it fails to accommodate several
phenomenological issues that remain as central questions in current
particle physics, such as the existence of non-zero neutrino
masses. This is nowadays an undeniable experimental fact due to the
measurements obtained by many neutrino oscillation experiments, which
have led us to an increasingly accurate knowledge of the relevant
parameters over the years \cite{deSalas:2017kay}.

The scientific literature contains a myriad of SM extensions with new
ingredients that generate neutrino masses. This includes models with
Dirac \cite{Ma:2016mwh,CentellesChulia:2018gwr} or Majorana neutrinos
\cite{Ma:1998dn}, with neutrino masses induced at tree-level or
radiatively \cite{Cai:2017jrq}, at low- \cite{Boucenna:2014zba} or
high-energy scales, and by operators with low or high dimensionalities
\cite{Anamiati:2018cuq}. One of the most common signatures of these
neutrino mass models is lepton flavor violation (LFV), which in many
scenarios may lead to observable rates in processes involving charged
leptons. This makes LFV a very important probe of neutrino mass
models, and more generally of models with extended lepton sectors. See
\cite{Calibbi:2017uvl} for a recent review on LFV.

Rare decays stand among the most powerful tests of the
SM. Interestingly, the LHCb collaboration has recently reported
several deviations between the measurements and the SM predictions in
observables associated to rare semileptonic B-meson decays involving a
$b \to s$ quark flavor transition. These include several angular
observables, the $P_5^\prime$ observable being the most popular one,
as well as the branching ratios of several processes, most notably
$B_s \to \phi \mu^+ \mu^-$ ~\cite{Aaij:2015oid,Aaij:2015esa}. Also
recently, the Belle collaboration presented an independent measurement
of $P_5^\prime$, compatible with the results obtained by
LHCb~\cite{Abdesselam:2016llu,Wehle:2016yoi}. In addition, the LHCb
collaboration has also measured the theoretically clean ratios
\begin{align}
R_{K^{(\ast)}} = \frac{ \int_{q^2_{\rm min}}^{q^2_{\rm max}} \frac{d\Gamma(B \rightarrow K^{(\ast)} \mu^+ \mu^-)}{dq^2} \, dq^2}{\int_{q^2_{\rm min}}^{q^2_{\rm max}} \frac{d\Gamma(B \rightarrow K^{(\ast)} e^+ e^-)}{dq^2} \, dq^2} \, ,
\end{align}
obtained for specific dilepton invariant mass squared ranges $q^2 \in
[q^2_{\rm min}, q^2_{\rm max}]$. In the SM, these ratios are expected
to be approximately equal to one, due to the fact that the SM gauge
bosons couple with the same strength to all three families of
leptons. These observables are precisely constructed to test this
feature of the SM, known as lepton flavor universality (LFU). It is
therefore very remarkable that LHCb reported values significantly
lower than one, in one $q^2$ bin of the $R_K$ ratio
\cite{Aaij:2014ora}, as well as in two $q^2$ bins of the $R_{K^\ast}$
ratio \cite{Aaij:2017vbb}:
\begin{align}
R_K &= 0.745^{+0.090}_{-0.074}\pm0.036    \,, \quad
q^2 \in [1,6]~\text{GeV}^2 \,, \nonumber \\[0.2cm]
R_{K^\ast} &= 0.660^{+0.110}_{-0.070}\pm0.024    \,, \quad
q^2 \in [0.045,1.1]~\text{GeV}^2\,, \nonumber \\[0.2cm]
R_{K^\ast} &= 0.685^{+0.113}_{-0.069}\pm0.047    \,, \quad
q^2 \in [1.1,6.0]~\text{GeV}^2 \,. 
\end{align}
These measurements imply deviations from the SM expected values
\cite{Descotes-Genon:2015uva,Bordone:2016gaq} at the $2.6\,\sigma$
level in the case of $R_K$, $2.2\,\sigma$ for $R_{K^\ast}$ in the
low-$q^2$ region, and $2.4\,\sigma$ for $R_{K^\ast}$ in the
central-$q^2$ region. Belle has also reported on the apparent
violation of LFU in the related observables $Q_4$ and $Q_5$
\cite{Wehle:2016yoi}. These observations and their potential New
Physics (NP) implications have made the \emph{$b \to s$ anomalies} a
subject of great interest.

It has been pointed out that the violation of lepton flavor
universality generically implies the violation of lepton flavor
\cite{Glashow:2014iga}. Although there are several explicit
counterexamples to this rule \cite{Celis:2015ara,Alonso:2015sja}, this
connection does indeed exist in most of the models introduced to
explain the $b \to s$ anomalies. In fact, this connection may be used
to learn about neutrino oscillation parameters
\cite{Boucenna:2015raa}. However, since many of these models do not
account for the observed neutrino masses and mixings, one may question
whether the most relevant LFV effects are generally induced by the
non-universal interactions associated to the $b \to s$ anomalies or by
the NP associated to the generation of neutrino
masses. Furthermore, even if the explanation to the $b \to s$
anomalies also involves LFV, the resulting rates could perhaps be too
low to be observed by the experiments taking place in the near
future. It is the goal of this paper to address these questions in a
particular model.

In this paper we consider a $Z^\prime$ model introduced to explain the
$b \to s$ anomalies \cite{Sierra:2015fma}, extended with a non-trival
\emph{embedding} of neutrino masses. As we will see below, the gauge
structure required to explain the $b \to s$ anomalies restricts the
model building for the generation of neutrino masses. Our focus will
be on the phenomenological exploration of the resulting LFV signatures
in this model, both at the usual low-energy experiments and in B-meson
decays. This has been studied previously for generic $Z^\prime$ models
in \cite{Crivellin:2015era,Becirevic:2016zri}.

The rest of the paper is organized as follows. In Sec. \ref{sec:btos}
we briefly review the current status of the $b \to s$ anomalies and
establish some basic notation to be used along the paper. In
Sec. \ref{sec:model} we introduce the model and discuss its most
relevant features. Our setup for the phenomenological analysis as well
as our results are described in detail in
Sec. \ref{sec:pheno}. Finally, we draw our conclusions in
Sec. \ref{sec:conclusions}.

\section{A brief review of the $\boldsymbol{b \to s}$ anomalies}
\label{sec:btos}

In order to interpret the available data on $b \to s$ transitions it
proves convenient to adopt an effective field theory language. The
effective Hamiltonian for $b \to s$ transitions is
\begin{equation} \label{eq:effH}
\mathcal H_{\text{eff}} = - \frac{4 G_F}{\sqrt{2}} \, V_{tb} V_{ts}^\ast \, \frac{e^2}{16 \pi^2} \, \sum_k \left(C_k \, \mathcal O_k + C^\prime_k \, \mathcal O^\prime_k \right) + \hc \, .
\end{equation}
Here $G_F$ is the Fermi constant, $e$ the electric charge and $V$ the
Cabibbo-Kobayashi-Maskawa (CKM) matrix. $\mathcal O_k$ and $\mathcal
O^\prime_k$ are the effective operators that contribute to $b \to s$
transitions, and $C_k$ and $C^\prime_k$ their Wilson coefficients.  It
is usually convenient to split the Wilson coefficients into the SM and
the NP contributions, $C_k = C_k^{\text{SM}} + C_k^{\text{NP}}$. In
the following we will indicate their leptonic flavor indices
explicitly. The operators that will be relevant for our discussion are
\begin{equation}
\mathcal O_9^{\ell_i \ell_j} = \left( \bar s \gamma_\mu P_L b \right) \, \left( \bar \ell_i \gamma^\mu \ell_j \right) \quad , \quad \mathcal O_{10}^{\ell_i\ell_j} = \left( \bar s \gamma_\mu P_L b \right) \, \left( \bar \ell_i \gamma^\mu \gamma_5 \ell_j \right) \, . \label{eq:ops}
\end{equation}
Primed operators are obtained by replacing $P_L$ by $P_R$ in the quark
current and $\ell_{i,j} = e, \mu, \tau$ are the three lepton
flavors. One can use data on $b \to s$ transitions to constrain the
Wilson coefficients of these operators. Interestingly, several
independent global fits
\cite{Capdevila:2017bsm,Altmannshofer:2017yso,DAmico:2017mtc,Hiller:2017bzc,Geng:2017svp,Ciuchini:2017mik,Alok:2017sui,Hurth:2017hxg}
have found that the tension between the SM predictions and the
experimental results can be alleviated with the introduction of a
negative NP contribution in $C_9^{\mu\mu}$, leading to a total Wilson
coefficient significantly smaller than the one in the SM. This has
driven a general interest in the $b \to s$ anomalies resulting in many
NP models aiming at an explanation of the experimental observations.

\section{The model}
\label{sec:model}

We consider an extended version of the model introduced in
\cite{Sierra:2015fma} that also accounts for the existence of non-zero
neutrino masses. A sketch of this version of the model was presented
in Sec. III.B of \cite{Sierra:2015fma}.

The gauge group of the model is $\mathrm{SU(3)_c} \times
\mathrm{SU(2)_L} \times \mathrm{U(1)_Y} \times \mathrm{U(1)_X}$, hence
extending the SM gauge symmetry with an additional $\mathrm{U(1)_X}$
factor. The gauge coupling associated to this symmetry will be denoted
by $g_X$ and the gauge boson by $Z^\prime$. Besides the usual SM
fields, neutral under $\mathrm{U(1)_X}$, the matter content of the
model is composed by one generation of vector-like (VL) quark doublets
$Q_{L,R} = \left( U , D \right)_{L,R}$, two generations of vector-like
lepton doublets $L_{L,R} = \left( N , E \right)_{L,R}$, the
electroweak singlet scalars $\phi$ and $S$ and two generations of
vector-like fermions $F$.~\footnote{The number of new fermion
  generations has been chosen following the principle of
  minimality. More generations are possible, but they are not required
  to accommodate the solar and atmospheric neutrino mass scales at
  tree-level.} All new fields are charged under $\mathrm{U(1)_X}$. The
complete scalar and fermion particle content of the model is given in
Table \ref{tab:particles}.

\begin{table}
\centering
\begin{tabular}{| c c c c c c |}
\hline  
 & generations & $\mathrm{SU(3)}_c$ & $\mathrm{SU(2)}_L$ & $\mathrm{U(1)}_Y$ & $\mathrm{U(1)}_X$ \\
\hline
\hline    
$H$ & 1 & ${\bf 1}$ & ${\bf 2}$ & $1/2$ & $0$ \\
$\phi$ & 1 & ${\bf 1}$ & ${\bf 1}$ & $0$ & $2$ \\
$S$ & 1 & ${\bf 1}$ & ${\bf 1}$ & $0$ & $-4$ \\ 
\hline
\hline    
$q_L$ & 3 & ${\bf 3}$ & ${\bf 2}$ & $1/6$ & $0$ \\   
$u_R$ & 3 & ${\bf 3}$ & ${\bf 1}$ & $2/3$ & $0$ \\    
$d_R$ & 3 & ${\bf 3}$ & ${\bf 1}$ & $-1/3$ & $0$ \\     
$\ell_L$ & 3 & ${\bf 1}$ & ${\bf 2}$ & $-1/2$ & $0$  \\     
$e_R$ & 3 & ${\bf 1}$ & ${\bf 1}$ & $-1$ & $0$ \\
$Q_{L,R}$ & 1 & ${\bf 3}$ & ${\bf 2}$ & $1/6$ & $2$ \\
$L_{L,R}$ & 2 & ${\bf 1}$ & ${\bf 2}$ & $-1/2$ & $2$ \\
$F_{L,R}$ & 2 & ${\bf 1}$ & ${\bf 1}$ & $0$ & $2$ \\
\hline
\hline
\end{tabular}
\caption{Scalar and fermion particle content of the model.}
\label{tab:particles}
\end{table}

The new Yukawa terms in the model are
\begin{equation} \label{eq:Yukawa}
- \mathcal L_Y = \lambda_Q \, \overline{Q_R} \, \phi \, q_L + \lambda_L \, \overline{L_R} \, \phi \, \ell_L + y \, \overline{L_L} \, H \, F_R + \widetilde y \, \overline{L_R} \, H \, F_L  + h \, S \, \overline{F_{L}^c} \, F_{L} + \widetilde h \, S \, \overline{F_{R}^c} \, F_{R} + \, \hc \, ,
\end{equation}
where $\lambda_L$ is a $2 \times 3$ matrix, $y$ and $\widetilde y$ are
$2 \times 2$ matrices and $h$ and $\widetilde h$ are $2 \times 2$
symmetric matrices. The $\lambda_Q$ and $\lambda_L$ couplings are the
only ones involving the SM fermions, and thus play a crucial role in
the resolution of the $b \to s$ anomalies. Furthermore, the
vector-like fermions $Q$, $L$ and $F$ have gauge invariant Dirac mass
terms
\begin{equation} \label{eq:Mass}
- \mathcal L_m = m_Q \, \overline{Q_L} Q_R + m_L \, \overline{L_L} L_R + m_F \, \overline{F_L} F_R + \hc \, .
\end{equation}
Both $m_L$ and $m_F$ are $2 \times 2$ matrices. The scalar potential
of the model can be split as
\begin{equation}
\mathcal V = \mathcal V_{\text{SM}} + \Delta \mathcal V \, .
\end{equation}
Here $\displaystyle \mathcal V_{\text{SM}} = m_H^2 |H|^2 +
\frac{\lambda}{2} |H|^4$ is the usual SM scalar potential. The new terms
involving the $\mathrm{U(1)_X}$ charged scalars are
\begin{align}
- \Delta \mathcal V = & \, m_\phi^2 \, |\phi|^2 + m_S^2 \, |S|^2 + \frac{\lambda_\phi}{2} \, |\phi|^4 + \frac{\lambda_S}{2} \, |S|^4 \nonumber \\
& + \lambda_{H \phi} \, |H|^2 |\phi|^2 + \lambda_{H S} \, |H|^2 |S|^2 + \lambda_{\phi S} \, |\phi|^2 |S|^2 + \left( \mu^\prime \, \phi^2 S + \hc \right) \, .
\end{align}
We will assume that the minimization of the potential leads to
non-zero vacuum expectation values (VEVs) for all scalars,
\begin{equation}
\langle H^0 \rangle = \frac{v}{\sqrt{2}} \, , \qquad \langle \phi \rangle = \frac{v_\phi}{\sqrt{2}} \, , \qquad \langle S \rangle = \frac{v_S}{\sqrt{2}} \, .
\end{equation}
Here $H^0$ is the neutral component of the SM Higgs doublet $H$. The
$\phi$ and $S$ fields will be responsible for the spontaneous breaking
of $\mathrm{U(1)_X}$, giving a mass to the $Z^\prime$,
\begin{equation} \label{eq:ZpMass}
m_{Z^\prime}^2 = 4 g_X^2 \left( v_\phi^2 + 4 v_S^2 \right) \, .
\end{equation}
In addition, $v_\phi$ will induce mixings between the vector-like
fermions and their SM counterparts thanks to the $\lambda_Q$ and
$\lambda_L$ Yukawa interactions in Eq. \eqref{eq:Yukawa}. As we will
show below, this mixing plays a crucial role in the phenomenology of
the model.

\subsection{Neutrino masses}
\label{subsec:numass}

\begin{figure}
\centering
\includegraphics[scale=0.5]{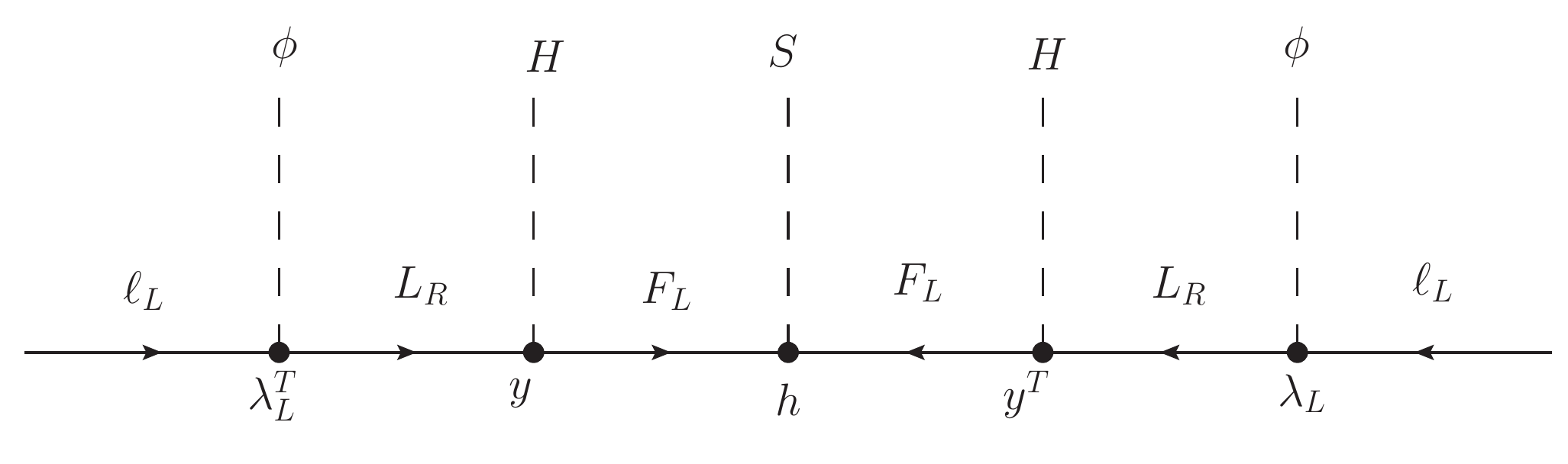}
\caption{Neutrino mass generation. We note that the model under
  discussion provides a specific ultraviolet completion to the
  dimension-8 operator $\displaystyle \mathcal O_\nu =
  \frac{1}{\Lambda_\nu^5} \ell \ell H H \phi \phi S$ pointed out in
  \cite{Sierra:2015fma}.}
\label{fig:numass}
\end{figure}

The definition of a conserved lepton number is not possible if $S$
gets a non-zero VEV. Indeed, $\displaystyle
\langle S \rangle = \frac{v_S}{\sqrt{2}} \ne 0$ breaks lepton number,
leading to Majorana neutrino masses~\footnote{Note, however, that
  lepton number conservation was actually enforced by the
  $\mathrm{U(1)_X}$ gauge symmetry. For instance, Majorana mass terms
  like $\overline{F_{L}^c} F_{L}$ were forbidden. For this reason, the
  spontaneous breaking of lepton number does not lead to the existence
  of a physical Goldstone boson, which is instead absorbed by the
  $Z^\prime$ boson.}. In order to find an expression for the light
neutrino masses, one must diagonalize the complete $11 \times 11$
neutral fermion mass matrix. In the basis $\mathcal N = \{ \nu_L , N_R^c ,
N_L , F_R^c , F_L \}$, this matrix takes the form
\begin{equation}
\mathcal M_{\mathcal N} = \left( \begin{array}{ccccccccccc}
0 & - \frac{1}{\sqrt{2}} v_\phi \lambda_L^T & 0 & 0 & 0 \\
- \frac{1}{\sqrt{2}} v_\phi \lambda_L & 0 & m_L^T & 0 & \frac{1}{\sqrt{2}} v \, \widetilde {y} \\
0 & m_L & 0 & - \frac{1}{\sqrt{2}} v \, y & 0 \\
0 & 0 & - \frac{1}{\sqrt{2}} v \, y^T & \sqrt{2} \, v_S \, \widetilde h & m_F^T \\
0 & \frac{1}{\sqrt{2}} v \, \widetilde{y}^T & 0 & m_F & \sqrt{2} \, v_S \, h
\end{array} \right) \, .
\end{equation}
The diagonalization of this matrix can be performed in \emph{seesaw
  approximation} by assuming $v_S \, h , v_S \, \widetilde h \ll v \,
y , v \, \widetilde y , v_\phi \, \lambda_L \ll m_{L,F}$. Importantly,
we note that in the absence of the Yukawa couplings $y$ and $h$,
$\widetilde y$ and $\widetilde h$ would not contribute to the
generation of neutrino masses at leading order, participating only at
higher orders in perturbation theory. For this reason, we will take
the simplifying assumption $\widetilde y = \widetilde h = 0$ in the
following. The resulting $3 \times 3$ mass matrix for the light
neutrinos is found to be
\begin{equation} \label{eq:mnumat}
m_\nu \simeq \frac{v^2 v_\phi^2 v_S}{2 \sqrt{2}} \, \lambda_L^T \, m_L^{-1} \, y \, m_F^{-1} \, h \, \left(m_F^{-1}\right)^T \, y^T \left( m_L^{-1} \right)^T \lambda_L  \, ,
\end{equation}
where higher order terms in $h \ll 1$ have been neglected. A
diagrammatic representation of the mechanism for neutrino mass
generation in this model is shown in Fig. \ref{fig:numass}.

A neutrino mass matrix as the one in Eq.~\eqref{eq:mnumat} formally
resembles that obtained in the inverse seesaw
\cite{Mohapatra:1986bd}. Indeed, neutrino masses get suppresed due to
the smallness of the $h v_S$ term, which allows for a low mass scale
for the states that participate in the generation of neutrino
masses. This justifies the choice $h \ll 1$, which is natural in the
sense of 't Hooft~\cite{tHooft:1979rat}, since the limit $h \to 0$
increases the symmetry of the model protecting this choice against
quantum corrections.~\footnote{We refer to \cite{Abada:2014vea} for a
  comprehensive exploration of possible inverse seesaw realizations.}

Given a specific texture for the $\lambda_L$ Yukawa matrices, one can
always find a matrix $h$ that reproduces the observed neutrino masses
and mixing angles. This matrix can be easily derived by inverting
Eq.~\eqref{eq:mnumat},
\begin{equation} \label{eq:hsol}
h = \bar v^{-5} \, m_F \, y^{-1} \, m_L \, \bar \lambda_L^T \, m_\nu \, \bar \lambda_L \, m_L^T \, \left(y^{-1}\right)^T \, m_F^T \, ,
\end{equation}
where $\bar \lambda_L$ is a $3 \times 2$ matrix such that $\lambda_L
\bar \lambda_L = \mathbb{I}_{2}$, $\mathbb{I}_{2}$ being the $2 \times
2$ unit matrix, and we have defined $\displaystyle \bar v^5 =
\frac{v^2 v_\phi^2 v_S}{2 \sqrt{2}}$. The neutrino mass matrix is
diagonalized as
\begin{equation}
U^{T} \, m_\nu \, U=\widehat{m}_\nu\equiv
\left(
\begin{array}{ccc}
m_1 & 0 & 0\\
0 & m_2 & 0\\
0 & 0 & m_3
\end{array}
\right) \, ,
\label{eq:mnudiag}
\end{equation}
where 
\begin{equation}
\label{eq:PMNS}
U=
\left(
\begin{array}{ccc}
 c_{12}c_{13} & s_{12}c_{13}  & s_{13}e^{i\delta}  \\
-s_{12}c_{23}-c_{12}s_{23}s_{13}e^{-i\delta}  & 
c_{12}c_{23}-s_{12}s_{23}s_{13}e^{-i\delta}  & s_{23}c_{13}  \\
s_{12}s_{23}-c_{12}c_{23}s_{13}e^{-i\delta}  & 
-c_{12}s_{23}-s_{12}c_{23}s_{13}e^{-i\delta}  & c_{23}c_{13}  
\end{array}
\right)
\end{equation}
is the standard leptonic mixing matrix. Here $\delta$ is the
CP-violating Dirac phase and we denote $c_{ij} = \cos \theta_{ij}$ and
$s_{ij} = \sin \theta_{ij}$.~\footnote{We note that the similarity to
  the usual inverse seesaw mass matrix would also allow one to use an
  adapted Casas-Ibarra parameterization \cite{Casas:2001sr}, as
  previously done in
  \cite{Basso:2012ew,Abada:2012mc,Abada:2014kba}. In this case, one
  solves Eq. \eqref{eq:mnumat} for the $\lambda_L$ matrix, obtaining
  the general expression $\lambda_L = \bar v^{-5/2} \, V_X^\dagger \,
  D_{\sqrt{X}} \, R \, D_{\sqrt{m_\nu}} \, U^\dagger$, where
  $D_{\sqrt{m_\nu}} = {\rm{diag}}(\sqrt{m_{\nu_i}})$, $D_{\sqrt{X}} =
  {\rm{diag}} (\sqrt{\hat X_i})$, with $\hat X_i$ the eigenvalues of
  $X = m_L^T \left(y^{-1}\right)^T m_F^T \, h^{-1} \, m_F \, y^{-1} \,
  m_L$, and $V_X$ is the matrix that diagonalizes $X$ as $V_X X V_X^T
  = \hat X$. $R$ is a $2 \times 3$ complex matrix such that $R R^T =
  \mathbb{I}_{2}$.}

\subsection{Solving the $\boldsymbol{b \to s}$ anomalies}
\label{subsec:btos}

\begin{figure}
\centering
\includegraphics[scale=0.5]{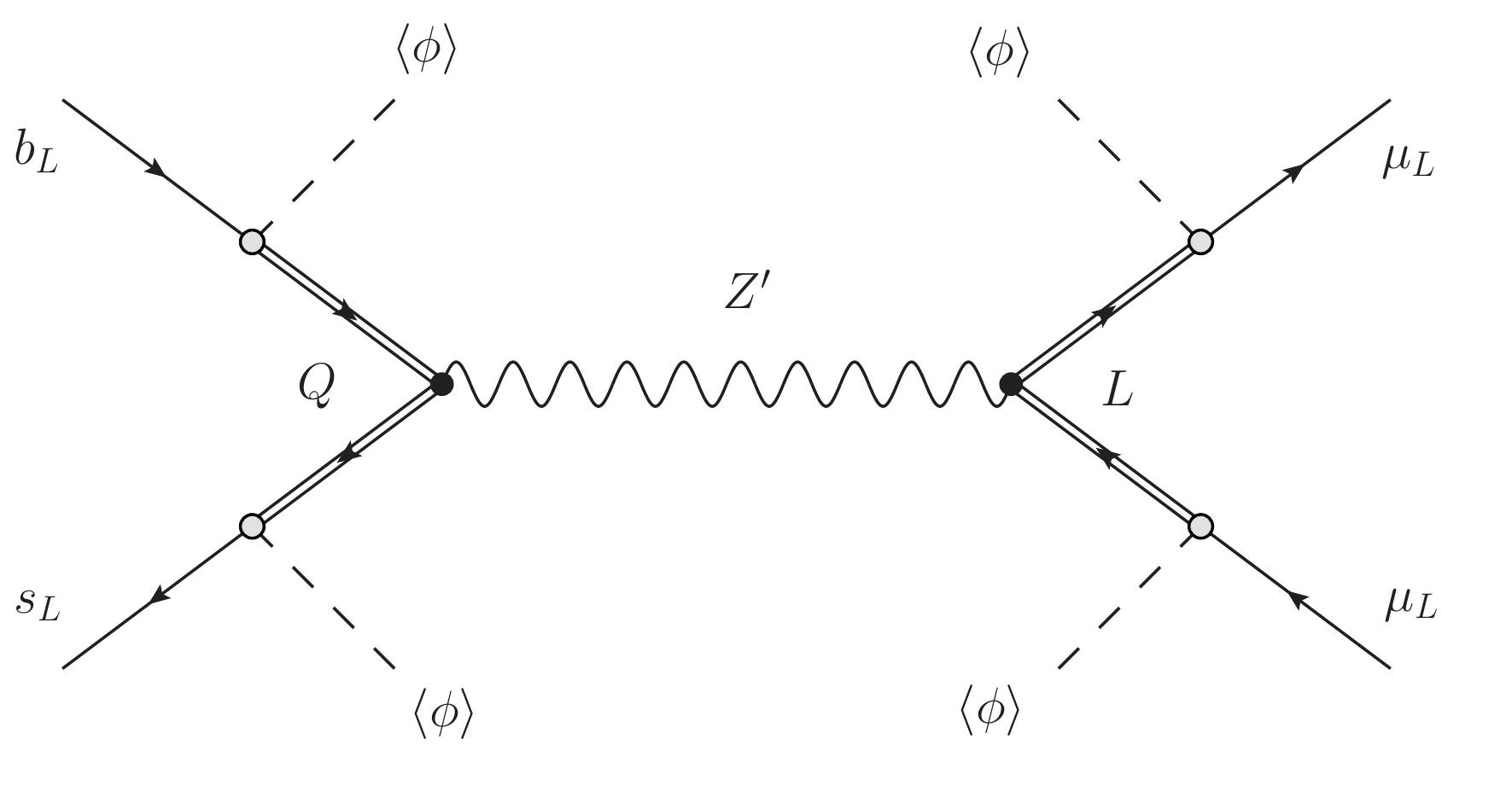}
\caption{Generation of $\mathcal O_9$ and $\mathcal O_{10}$. The
  mixing between the SM fermions and the VL ones induce semileptonic
  four-fermion interactions.}
\label{fig:couplings}
\end{figure}

The solution to the $b \to s$ anomalies follows the same lines as in
\cite{Sierra:2015fma}. The spontaneous breaking of the
$\mathrm{U(1)_X}$ gauge symmetry by the $\phi$ VEV induces mixings
between the SM and VL fermions due to the $\lambda_Q$ and $\lambda_L$
Yukawa couplings. Defining the bases $\mathcal D_{L,R} = \left\{ d , D
\right\}_{L,R}$ and $\mathcal E_{L,R} = \left\{ e , E \right\}_{L,R}$,
the Lagrangian after symmetry breaking includes the terms
\begin{equation}
- \mathcal L \supset \overline{\mathcal D_L} \mathcal M_{\mathcal D} \mathcal D_R + \overline{\mathcal E_L} \mathcal M_{\mathcal E} \mathcal E_R + \hc \, .
\end{equation}
The $4 \times 4$ down-quark mass matrix is given by
\begin{equation} \label{eq:Md}
\mathcal{M}_{\mathcal D} = \left( \begin{array}{cc}
\frac{1}{\sqrt{2}} v Y_d & \frac{1}{\sqrt{2}} v_\phi \lambda_Q^T \\
0 & -m_Q \end{array} \right) \, ,
\end{equation}
whereas the $5 \times 5$ charged lepton mass matrix is
\begin{equation} \label{eq:Me}
\mathcal{M}_{\mathcal E} = \left( \begin{array}{ccc}
\frac{1}{\sqrt{2}} v Y_e & \frac{1}{\sqrt{2}} v_\phi \lambda_L^T \\
0 & -m_L \end{array} \right) \, ,
\end{equation}
with the SM Yukawa couplings defined as $Y_d \, H \, \overline{q_L} \,
d_R$ and $Y_e \, H \, \overline{\ell_L} \, e_R$. These two fermion
mass matrices can be diagonalized by means of the following biunitary
transformations
\begin{align}
\mathcal D_L = V_d \, \widehat{\mathcal D_L} \, , \quad & \quad \mathcal D_R = U_d \, \widehat{\mathcal D_R} \, , \\
\mathcal E_L = V_e \, \widehat{\mathcal E_L} \, , \quad & \quad \mathcal E_R = U_e \, \widehat{\mathcal E_R} \, ,
\end{align}
where $V_{d,e}$ and $U_{d,e}$ are unitary matrices and
$\widehat{\mathcal D_{L,R}}$ and $\widehat{\mathcal E_{L,R}}$ denote
the physical mass eigenstates. With these definitions, the diagonal
mass matrices $\widehat{\mathcal M_{\mathcal D}}$ and
$\widehat{\mathcal M_{\mathcal E}}$ are obtained as $\widehat{\mathcal
  M_{\mathcal D}} = V_d^\dagger \, \mathcal M_{\mathcal D} \, U_d$ and
$\widehat{\mathcal M_{\mathcal E}} = V_e^\dagger \, \mathcal
M_{\mathcal E} \, U_e$, respectively.

The SM-VL mixing leads to the generation of $Z^\prime$ effective
couplings to the SM fermions. If these are parametrized as
\cite{Buras:2012jb,Altmannshofer:2014rta}
\begin{equation}
\mathcal L \supset \bar f_i \gamma^\mu \left( \Delta_L^{f_i f_j} P_L + \Delta_R^{f_i f_j} P_R \right) f_j Z_\mu^\prime \, .
\end{equation}
the $Z^\prime-b-s$ and $Z^\prime-\mu-\mu$ couplings, relevant for the
explanation of the $b \to s$ anomalies, are given by
\begin{align}
\Delta_L^{bs} & = - 2 \, g_X \, \left(V_d\right)_{42}^\ast \, \left(V_d\right)_{43} \, , \label{eq:Deltabs} \\
\Delta_L^{\mu \mu} & = - 2 \, g_X \, \sum_{k=4,5} \left(V_e\right)_{k2}^\ast \, \left(V_e\right)_{k2} \, . \label{eq:Deltamumu}
\end{align}
These couplings lead to a tree-level contribution to the four-fermion
operators $\mathcal O_9$ and $\mathcal O_{10}$, as shown in
Fig. \ref{fig:couplings}. In fact, since the SM fermions participating
in the effective vertices are purely left-handed, the operators
$\mathcal O_9$ and $\mathcal O_{10}$ are generated simultaneously,
with their Wilson coefficients fulfilling~\cite{Altmannshofer:2014rta}
\begin{equation} \label{eq:c9c10}
C_9^{\mu\mu , \text{NP}} = - C_{10}^{\mu\mu , \text{NP}} = - \frac{\Delta_L^{bs} \Delta_L^{\mu \mu}}{V_{tb} V_{ts}^\ast} \, \left( \frac{\Lambda_v}{m_{Z^\prime}} \right)^2 \, ,
\end{equation}
where we have defined
\begin{equation} \label{eq:Lambdav}
\Lambda_v = \left( \frac{\pi}{\sqrt{2} G_F \alpha} \right)^{1/2} \simeq 4.94 \, \text{TeV} \, ,
\end{equation}
with $\alpha = \frac{e^2}{4 \pi}$ the electromagnetic fine structure
constant. With these ingredients at hand, it is straightforward to
check that the model under discussion can reproduce the required value
for $C_9^{\mu\mu , \text{NP}}$ found by the global fits to $b \to s$
data. In our numerical analysis we will always consider parameter
values that do so. Furthermore, analogous operators with violation of
lepton flavor are also induced. Generalizing Eq. \eqref{eq:Deltamumu}
to
\begin{equation} \label{eq:Deltalilj}
\Delta_L^{\ell_i \ell_j} = - 2 \, g_X \, \sum_{k=4,5} \left(V_e\right)_{ki}^\ast \, \left(V_e\right)_{kj} \, ,
\end{equation}
one also has
\begin{equation} \label{eq:c9lilj}
C_9^{\ell_i \ell_j , \text{NP}} = - \frac{\Delta_L^{bs} \Delta_L^{\ell_i \ell_j}}{V_{tb} V_{ts}^\ast} \, \left( \frac{\Lambda_v}{m_{Z^\prime}} \right)^2 \, .
\end{equation}
The $C_9^{\ell_i \ell_j , \text{NP}}$ LFV Wilson coefficients are the
source of the B-meson LFV decays discussed in this work.

\subsection{Dark matter}
\label{subsec:DM}

Finally, we note that the setup described here can be minimally
extended to account for the dark matter of the Universe. Indeed, the
original model introduced in \cite{Sierra:2015fma} was the first NP
model addressing the $b \to s$ anomalies with a dark sector. This was
accomplished by adding the complex scalar $\chi$, with charges $\left(
{\bf 1}, {\bf 1}, 0 , -1 \right)$ under $\mathrm{SU(3)_c} \times
\mathrm{SU(2)_L} \times \mathrm{U(1)_Y} \times
\mathrm{U(1)_X}$. Assuming that this scalar does not get a VEV, the
breaking of the $\mathrm{U(1)_X}$ gauge symmetry leaves a remnant
$\mathbb{Z}_2$ parity, under which $\chi$ is odd. This mechanism
\cite{Krauss:1988zc,Petersen:2009ip,Sierra:2014kua} automatically
stabilizes $\chi$ and makes it a valid dark matter
candidate. Furthermore, the heavy $Z^\prime$ boson, crucial for the
explanation of the $b \to s$ anomalies, serves as a portal between the
SM and dark sectors. This establishes a non-trivial link between these
two phenomenological directions in the model. We refer to
\cite{Sierra:2015fma} for a detailed discussion of the dark matter
phenomenology of the model and to \cite{Vicente:2018xbv} for a recent
review on the possible connection between the $b \to s$ anomalies and
the dark matter of the Universe.

\section{Phenomenological analysis}
\label{sec:pheno}

Our phenomenological analysis uses the {\tt FlavorKit}
\cite{Porod:2014xia} functionality of {\tt SARAH}
\cite{Staub:2008uz,Staub:2009bi,Staub:2010jh,Staub:2012pb,Staub:2013tta}
for the analytical computation of the purely leptonic LFV
observables.~\footnote{For a pedagogical introduction to {\tt SARAH}
  in the context of non-supersymmetric models see
  \cite{Vicente:2015zba}.} This allows us to automatically obtain
complete analytical results for the LFV observables as well as robust
numerical routines when this is used in combination with {\tt SPheno}
\cite{Porod:2003um,Porod:2011nf}. For the calculation of the $B$-meson
LFV branching ratios we follow \cite{Crivellin:2015era}.

Let us now explain our parameter choices. Without loss of generality,
the matrices $m_L$ and $m_F$ will be taken to be diagonal. We will
also further assume a diagonal form for the $y$ matrix. Regarding the
fit to neutrino oscillation data, we will consider a specific
structure for the $\lambda_L$ matrix with $\left(\lambda_L\right)_{i1}
= 0$, thus forcing the matrix $h$ to contain flavor-violating
entries. The matrix $h$ will be obtained by using
Eq. \eqref{eq:hsol}.~\footnote{One could also consider an alternative
  scenario with $h = \bar{h} \, \mathbb{I}_{3}$, so that the only
  source of flavor violation is the matrix $\lambda_L$. However, such
  a general $\lambda_L$ matrix would potentially lead to $C_9^{e e ,
    \text{NP}} \ne 0$ and non-zero $\mu-e$ flavor violating
  amplitudes, making this scenario a very constrained one. We found
  that in order to avoid the stringent limits derived from flavor and,
  simultaneously, be compatible with neutrino oscillation data, a
  strong fine-tuning would be required. For this reason, we have not
  explored this scenario any further.} Finally, we make the choice
$\left(\lambda_Q\right)_{1} = 0$ in order to suppress the $Z^\prime$
couplings to 1st generation quarks.

In what concerns the parameter ranges explored in the following
analysis, we must take into account constraints derived from direct
searches at the Large Hadron Collider (LHC). These include searches
for the vector-like fermions in the model, as well as for the heavy
$Z^\prime$ boson that mediates the NP contributions to the flavor
observables. Regarding the $Z^\prime$ boson, one may naively think
that its production cross-section would be too low to be observable at
the LHC due to our choice $\left(\lambda_Q\right)_{1} = 0$. However,
the $Z^\prime$ can indeed be produced in $pp$ collisions due to the
non-vanishing heavy quark content in the protons. Due to the large
couplings to muons required to explain the $b \to s$ anomalies, it is
expected to decay mainly into $\mu^+ \mu^-$ (and, optionally, $\tau^+
\tau^-$ if the $\left(\lambda_L\right)_{i3}$ couplings take large
values). ATLAS \cite{Aaboud:2017buh} and CMS \cite{Sirunyan:2018exx}
have searched for a $Z^\prime$ boson in the dimuon channel but the
resulting limits are not very stringent, allowing for $Z^\prime$
masses as low as $\sim 100$ GeV, see \cite{Xu:2018} for a recent
analysis. Ditau searches are more sensitive and require $m_{Z^\prime}
\gtrsim 1$ TeV unless the $Z^\prime$ has a very large decay width
\cite{Faroughy:2016osc}. However, in our setup the $Z^\prime \to
\tau^+ \tau^-$ branching ratio will never be dominant due to the large
couplings to muons, and hence $m_{Z^\prime} \sim 1$ TeV will be
perfectly allowed. The LHC collaborations have also searched for the
vector-like fermions in the model, which provide complementary
collider bounds. The vector-like quarks are colored particles and thus
efficiently produced via QCD interactions at the LHC. This implies
lower bounds on their mass slightly above the TeV scale
\cite{Cacciapaglia:2018lld}. Since our setup works with vector-like
quark masses above this scale, the existing bounds can be easily
satisfied. Finally, the vector-like leptons can also be searched for
in multilepton final states. The current limits are weaker than those
for vector-like quarks and allow for masses below the TeV
\cite{Xu:2018}. These constraints will be taken into account in the
numerical analysis that follows.

We now proceed to present the main numerical results of our analysis.

\subsection{BR($\boldsymbol{B \to K \tau \mu}$) vs BR($\boldsymbol{\tau \to 3 \, \mu}$)}

We first discuss the correlation between $\text{BR}(B \to K \tau \mu)$
and $\text{BR}(\tau \to 3 \, \mu)$ and how it can be used to estimate
an upper bound for $\text{BR}(B \to K \tau \mu)$.~\footnote{See
  \cite{Guadagnoli:2018ojc} for a scenario leading to correlations
  between $\text{BR}(B \to K \tau \mu)$ and $\text{BR}(\tau \to 3 \,
  \mu)$ and $R_K$.} Assuming that the dominant contributions are
induced by the tree-level exchange of the $Z^\prime$ boson (see below
for a discussion on this point), the branching ratios for the $B \to K
\tau \mu$ and $\tau \to 3 \, \mu$ decays can be written as
\cite{Crivellin:2015era}
\begin{align}
\text{BR}(B \to K \tau \mu) & = \text{BR}(B \to K \tau^- \mu^+) + \text{BR}(B \to K \tau^+ \mu^-) = \nonumber \\
& = 2 \cdot 10^{-9} \, A_{K\tau\mu} \, \left| \frac{\Delta_L^{bs} \Delta_L^{\tau \mu}}{V_{tb} V_{ts}^\ast} \right|^2 \left( \frac{\Lambda_v}{m_{Z^\prime}} \right)^4 \, , \label{eq:BRBtoKtaumu} \\
\text{BR}(\tau \to 3 \, \mu) & = \frac{m_\tau^5}{768 \pi^3 \Gamma_\tau m_{Z^\prime}^4} \left| \Delta_L^{\mu \mu} \Delta_L^{\tau \mu} \right|^2 \, , \label{eq:BRtau3mu}
\end{align}
where $m_\tau$ and $\Gamma_\tau$ are the tau lepton mass and decay
width, respectively, and $A_{K\tau\mu} = 19.6 \pm 1.7$. This parameter
has been obtained by combining the coefficients $a_{K\tau\mu} +
b_{K\tau\mu}$, see \cite{Crivellin:2015era}, and adding the
$a_{K\tau\mu}$ and $b_{K\tau\mu}$ errors in quadrature.  We note that
although Ref. \cite{Becirevic:2016zri} provides slightly different
numerical values for these coefficients, they are perfectly
compatible, in particular given the level of precision required for
our analysis. One can now combine these expressions with
Eq. \eqref{eq:c9c10} to obtain
\begin{equation} \label{eq:corr}
\frac{\text{BR}(B \to K \tau \mu)}{\text{BR}(\tau \to 3 \, \mu)} = 1.7 \cdot 10^7 \, \text{TeV}^4 \left( \frac{\left| \Delta_L^{bs} \right|}{m_{Z^\prime}} \right)^4 \, \frac{1}{\left|C_9^{\mu\mu , \text{NP}}\right|^2} \, .
\end{equation}
The ratio $\left| \Delta_L^{bs} \right| / m_{Z^\prime}$ is strongly
constrained by $B_s - \overline{B_s}$ mixing, which in this model
would be induced via $Z^\prime$ tree-level exchange. Allowing for a
$10 \%$ deviation in the mixing amplitude, one finds
\cite{Altmannshofer:2014rta}~\footnote{The impact of stronger $B_s -
  \overline{B_s}$ mixing bounds has been recently explored in
  \cite{DiLuzio:2017fdq}.}
\begin{equation}
\frac{m_{Z^\prime}}{\left| \Delta_L^{bs} \right|} \gtrsim 244 \, \text{TeV} \quad \Rightarrow \quad \frac{\left| \Delta_L^{bs} \right|}{m_{Z^\prime}} \lesssim 4 \cdot 10^{-3} \, \text{TeV}^{-1} \, .
\end{equation}
Furthermore, the current experimental upper bound on $\text{BR}(\tau
\to 3 \, \mu)$ has been set by the Belle collaboration, which obtained
$\text{BR}(\tau \to 3 \, \mu)_{\text{max}} =
2.1\times10^{-8}$~\cite{Hayasaka:2010np}, whereas the preferred $2
\sigma$ range obtained for $C_9^{\mu\mu , \text{NP}}$ in the global
fit \cite{Capdevila:2017bsm} is $\left[ -0.88 , -0.37 \right]$. With
these ingredients at hand one can easily obtain the largest branching
ratio for the $B \to K \tau \mu$ decay in this model, finding
\begin{equation} \label{eq:maxBR}
\text{BR}(B \to K \tau \mu)_{\text{max}} \lesssim 8 \cdot 10^{-10} \, .
\end{equation}
This result is clearly below the current experimental limit,
$\text{BR}(B \to K \tau \mu) < 4.8 \cdot 10^{-5}$
\cite{Amhis:2014hma}. The main reason behind this result is the
stringent constraint from $B_s - \overline{B_s}$ mixing. However, we
would like to emphasize two points: (1) this is the largest
$\text{BR}(B \to K \tau \mu)$ that one expects when the $Z^\prime$
boson has purely left-handed couplings, as in the model under
consideration, and (2) while in models with additional $Z^\prime$
right-handed couplings cancellations in the $B_s - \overline{B_s}$
mixing amplitude are possible~\cite{Crivellin:2015era}, increasing
$\text{BR}(B \to K \tau \mu)_{\text{max}}$ beyond the value given in
Eq. \eqref{eq:maxBR} would require a significant fine-tuning of the
parameters.

\begin{figure}
\centering
\includegraphics[scale=0.7]{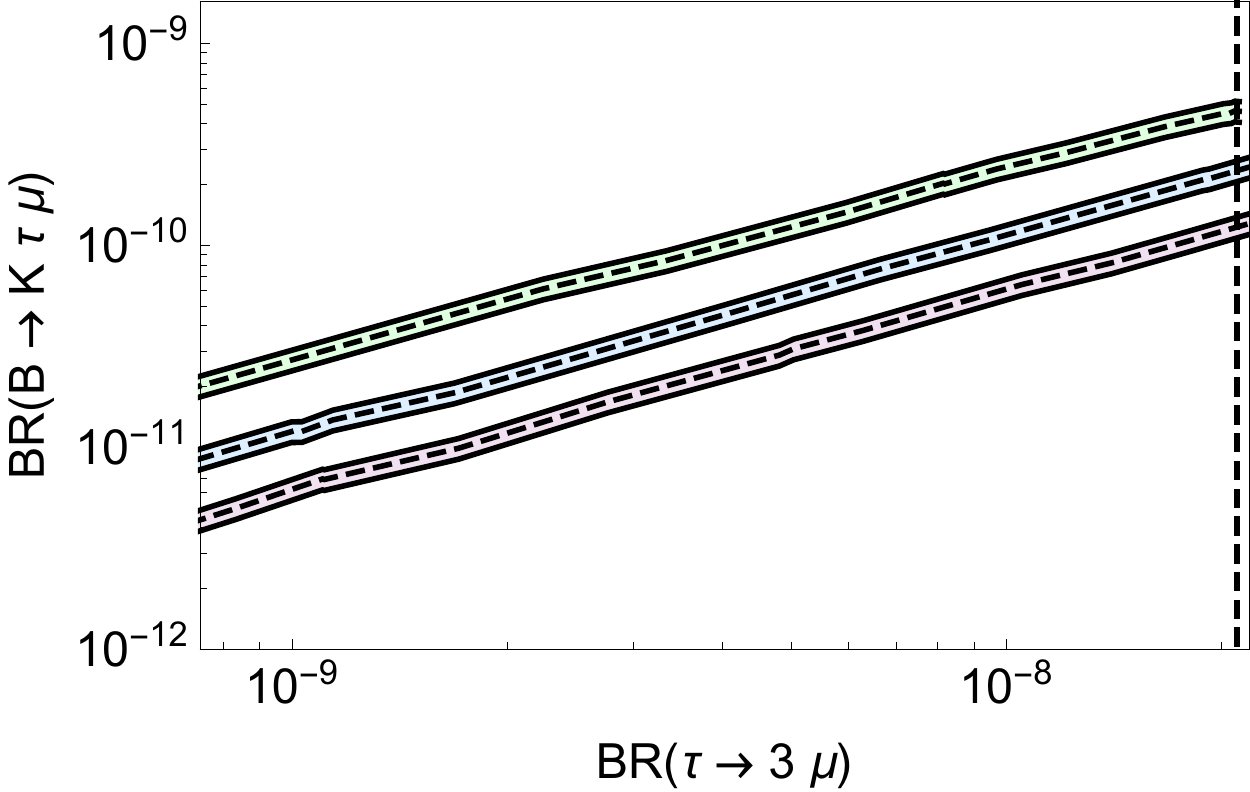}
\caption{Correlation between BR($\tau \to 3\mu$) and BR($B \to K \tau
  \mu$) for three different sets of parameters. This figure has been
  obtained varying $\left(\lambda_L\right)_{13} =
  \left(\lambda_L\right)_{22}$.  The vertical dashed line corresponds
  to the Belle experimental bound BR($\tau \to
  3\,\mu$)$_{\text{max}}=2.1 \cdot 10^{-8}$~\cite{Hayasaka:2010np}.}
\label{fig:correlation}
\end{figure}

Figure \ref{fig:correlation} shows the correlation between
$\text{BR}(B \to K \tau \mu)$ and $\text{BR}(\tau \to 3 \, \mu)$ for
three specific parameter choices. This figure has been obtained
varying $\left(\lambda_L\right)_{13} =
\left(\lambda_L\right)_{22}$. The values of the model parameters in
the three different scenarios are:

\begin{itemize}
\item {\bf Green:} $g_X=0.155$, $v_S=10.6$ GeV, $m_{Z^\prime}=1592$
  GeV, $\left(m_L\right)_{11}=\left(m_L\right)_{22}=1904$ GeV and
  $\left(\lambda_{Q}\right)_2=\left(\lambda_{Q}\right)_3=0.0407$.
\item {\bf Blue:} $g_X=0.2$, $v_S=200$ GeV, $m_{Z^\prime}=1010$ GeV,
  $\left(m_L\right)_{11}=\left(m_L\right)_{22}=1600$ GeV,
  $\left(\lambda_{Q}\right)_2=\left(\lambda_{Q}\right)_3=0.055$.
\item {\bf Purple:} $g_X=0.4$, $v_S=34$ GeV, $m_{Z^\prime}=2330$ GeV,
  $\left(m_L\right)_{11}=\left(m_L\right)_{22}=1007$ GeV,
  $\left(\lambda_{Q}\right)_2=\left(\lambda_{Q}\right)_3=0.052$.
\end{itemize}

We note that higher values of
$\left(\lambda_{Q}\right)_2=\left(\lambda_{Q}\right)_3$ would be
excluded due to $B_s - \overline{B_s}$ mixing constraints. The green
band in Fig. \ref{fig:correlation} reaches $\text{BR}(B \to K \tau
\mu) \sim 6 \cdot 10^{-10}$, close to the upper bound estimated in
Eq. \eqref{eq:maxBR}. As we will show next, the strong correlations
found in our analysis can be broken by loop effects, hence affecting
the general conclusions derived from our phenomenological
exploration. For instance, in regions of parameter space where loop
corrections cancel the tree-level results for $\text{BR}(\tau \to 3 \,
\mu)$, Eq. \eqref{eq:corr} would no longer hold and a larger
$\text{BR}(B \to K \tau \mu)$ would be allowed. This would require a
fine-tuning of the masses and mixings in the charged lepton sector.

\subsection{On the relevance of loop effects in BR($\boldsymbol{\tau \to 3 \, \mu}$)}

\begin{figure}
\centering
\includegraphics[scale=0.7]{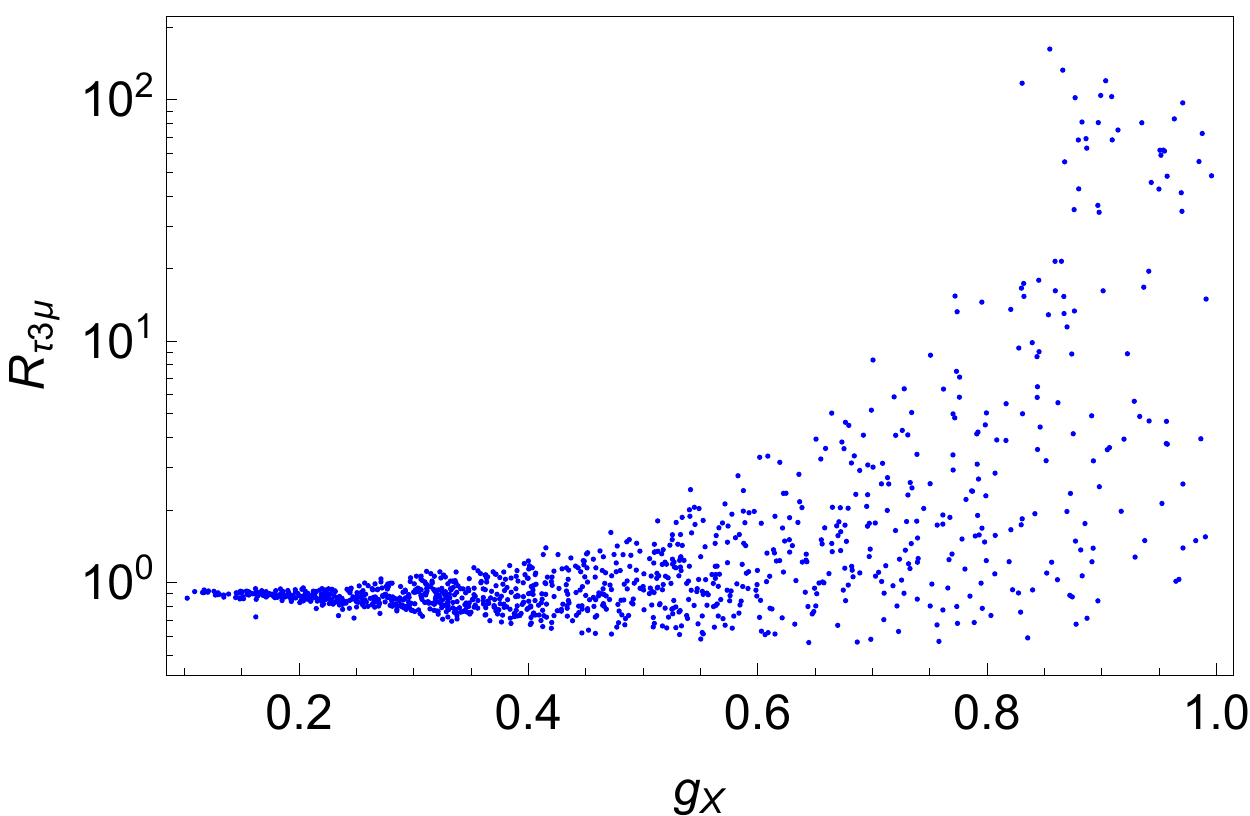}
\caption{Behavior of the ratio $R_{\tau 3 \mu}$ as a function of the
  gauge coupling $g_X$. Several model parameters have been randomly
  scanned over a wide range of numerical values, see text for
  details. The tree-level expression in Eq. \eqref{eq:BRtau3mu} and
  the complete numerical result including 1-loop corrections can be
  very different for $g_X \gtrsim 0.4$.}
\label{fig:loopeff}
\end{figure}

So far we have discussed tree-level predictions of the model. However,
one may wonder whether loop corrections might alter the results
presented above. We have addressed this issue in
Fig. \ref{fig:loopeff}, where we show the ratio between the tree-level
expression for BR($\tau \to 3\,\mu$) given in Eq. \eqref{eq:BRtau3mu}
and the complete numerical result including 1-loop contributions as
returned by {\tt SPheno},
\begin{equation}
R_{\tau 3 \mu} = \frac{\text{BR}(\tau \to 3 \, \mu)_{\text{tree-level}}}{\text{BR}(\tau \to 3 \, \mu)_{\text{1-loop}}} \, .
\end{equation}
This plot has been obtained by randomly scanning in the following
ranges:
\begin{eqnarray*}
 0.05&<&g_X<1.0\\
 10 \text{ GeV}&<&v_S<500 \text{ GeV}\\
 0.01&<&\left(\lambda_{Q}\right)_2=\left(\lambda_{Q}\right)_3<0.1\\
 0.8 \text{ TeV}&<& \left(m_L\right)_{11} = \left(m_L\right)_{22}<2 \text{ TeV}\\
 1 \text{ TeV}&<& m_{Z^\prime}<3 \text{ TeV}
\end{eqnarray*}
One can clearly see in Fig. \ref{fig:loopeff} that while the
tree-level expression in Eq. \eqref{eq:BRtau3mu} and the complete
numerical result including 1-loop corrections are actually very
similar for low values of $g_X$, they can be very different for
$g_X>0.4$.

\begin{figure}
\centering
\includegraphics[scale=0.45]{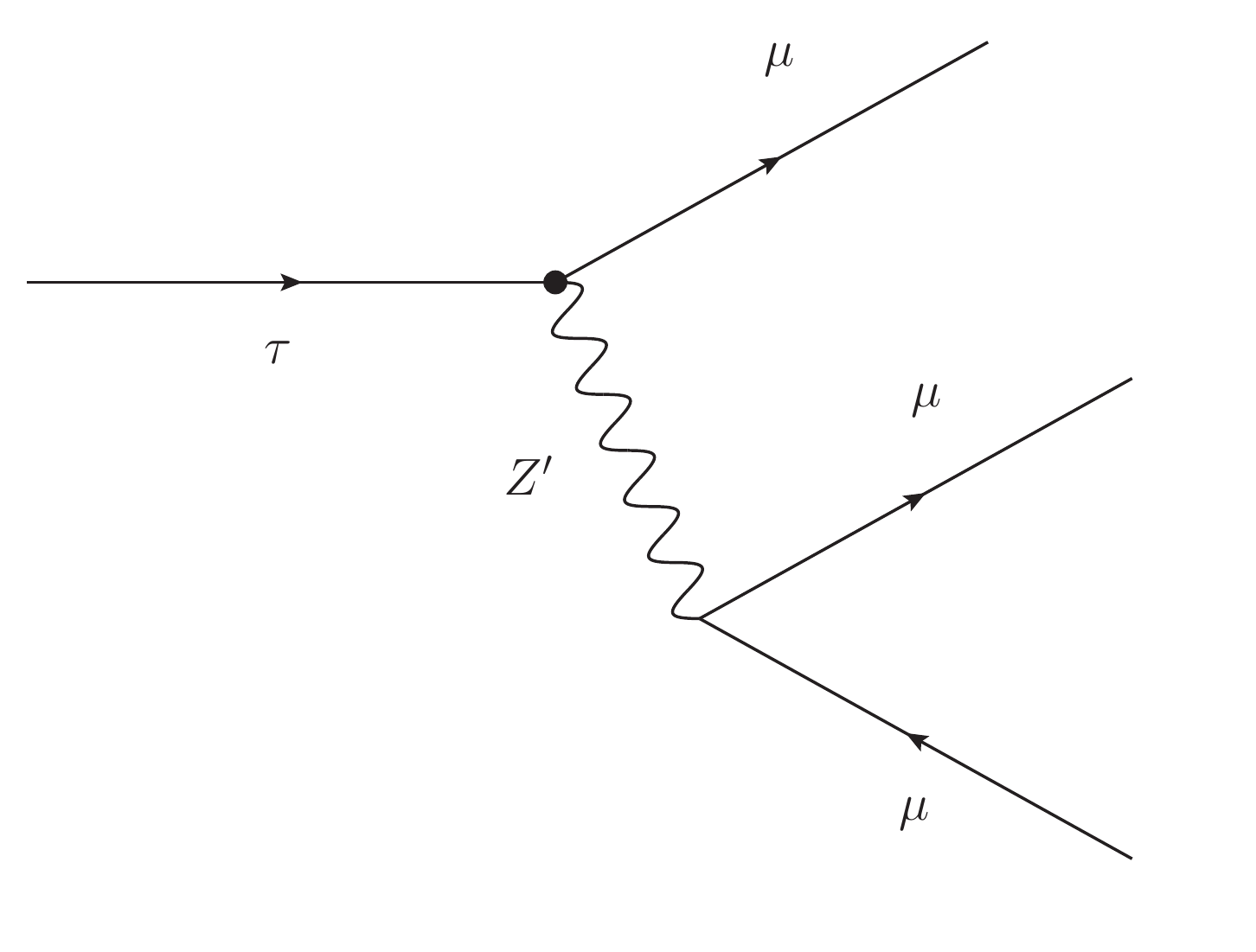}
\includegraphics[scale=0.45]{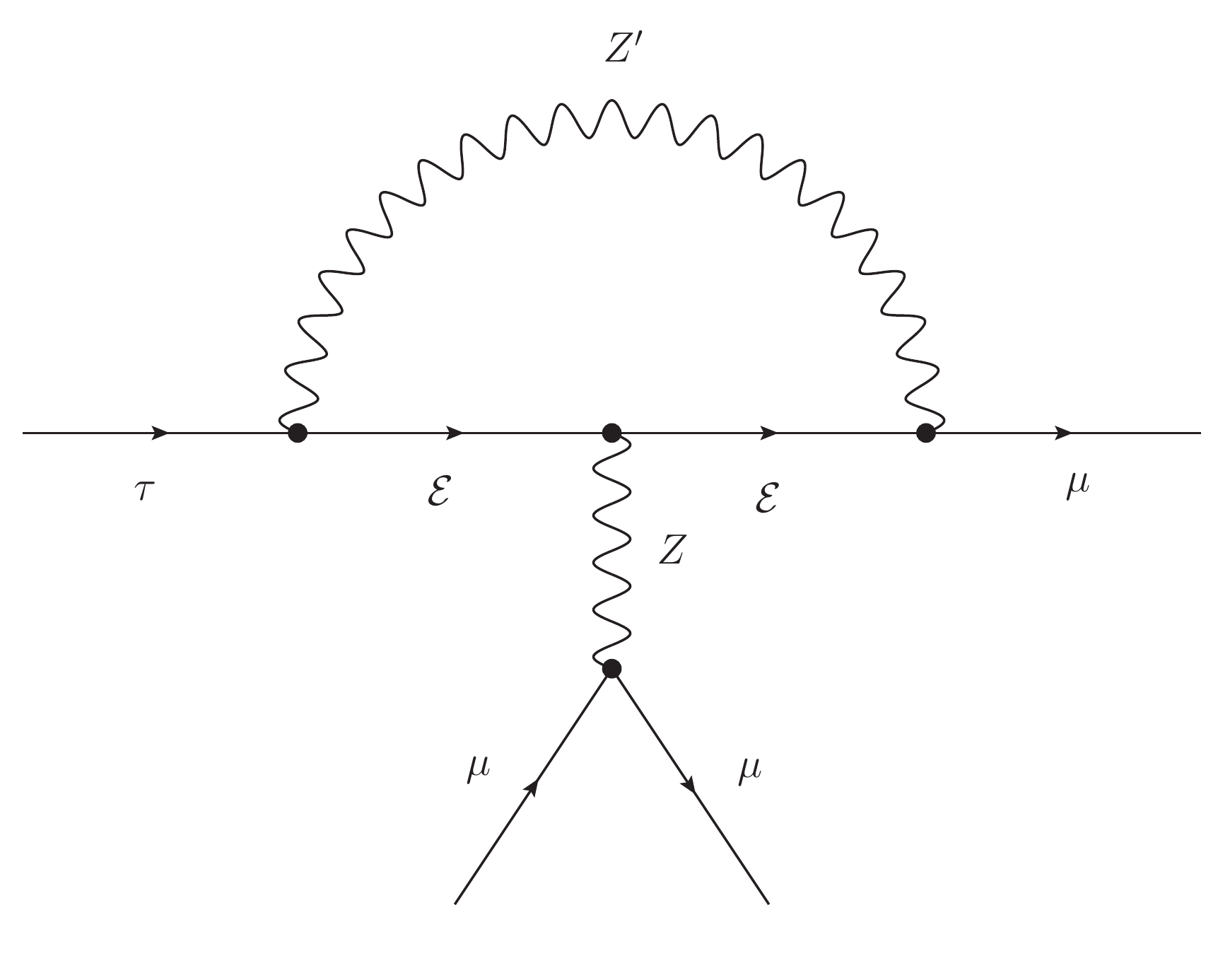}
\caption{Feynman diagrams relevant for the calculation of the $\tau
  \to 3\,\mu$ amplitude. On the left, the dominant tree-level
  contribution is shown, whereas the diagram on the right is one of
  the dominant 1-loop contributions. We note that the 1-loop diagram
  on the right should be accompanied by two diagrams with the $Z$
  boson line attached to one of the external lepton legs.}
\label{fig:loopeff-diag}
\end{figure}

The impact of the loop corrections in $\tau \to 3\,\mu$ can be easily
understood with the following considerations. In fact, it is not
surprising that loop effects can be as large as the tree-level ones in
$\tau \to 3\,\mu$. Fig. \ref{fig:loopeff-diag} shows two Feynman
diagrams relevant for the calculation of the $\tau \to 3\,\mu$
amplitude. The diagram on the left constitutes the dominant tree-level
contribution, whereas the diagram on the right is one of the dominant
1-loop contributions. Their contribution to the amplitude for external
left-handed leptons can be generically written as
\begin{align}
\mathcal A_{\rm tree} &= \frac{g_X^2}{m_{Z^\prime}^2} \, F_{\rm tree} \left( V_e \right) \, , \\
\mathcal A_{\rm loop} &= \frac{1}{16 \pi^2} \, \frac{g_X^2 \, g_{Z\ell\ell}}{m_{Z}^2} \, F_{\rm loop}^{g_X} \left( m_{\mathcal E} , V_e \right) \, ,
\end{align}
where $g_{Z\ell\ell}$ is the SM $Z$ boson coupling to a pair of
left-handed charged leptons and $F_{\rm tree}$ and $F_{\rm
  loop}^{g_X}$ are two functions of the charged leptons (the five
eigenstates) masses and mixings. $F_{\rm tree}$ only depends on the
mixings in $V_e$ due the $Z^\prime$ couplings to $\tau \mu$ and $\mu
\mu$, given in Eq. \eqref{eq:Deltalilj}. In contrast, $F_{\rm
  loop}^{g_X}$ also depends on the five charged lepton masses,
$m_{\mathcal E}$, due to the corresponding loop function. We first
note that for $F_{\rm tree} \simeq F_{\rm loop}^{g_X}$, both
contributions have comparable sizes, since
\begin{equation}
\frac{1}{16 \pi^2} \, \frac{1}{m_{Z}^2} \sim \frac{1}{m_{Z^\prime}^2} \quad , \quad \text{for} \, m_{Z^\prime} \sim \text{TeV} \, .
\end{equation}
Therefore, one would naively expect that loop effects in $\tau \to 3
\, \mu$ will be generically of a size that is comparable to the
tree-level ones. This is indeed what we find for large values of
$g_X$. Moreover, we note that the 1-loop contributions may have a
relative sign with respect to the tree-level ones, thus leading to
cancellations in the final amplitudes, as shown in
Fig. \ref{fig:loopeff}. In contrast, this is not the case for low
values of $g_X$ ($g_X \lesssim 0.4$). In this region of the parameter
space we find that $F_{\rm loop}^{g_X}$ is strongly reduced, hence
suppressing loop contributions. This is due to the fact that, although
$F_{\rm loop}^{g_X}$ does not depend explicitly on $g_X$, there is an
indirect dependence on this gauge coupling. In order to keep
$m_{Z^\prime}$ in the TeV ballpark for low values of $g_X$, one must
introduce a large $v_\phi$ VEV, see Eq. \eqref{eq:ZpMass}, and this in
turn affects the charged lepton masses and mixings as shown in
Eq. \eqref{eq:Me}. We have checked in detail that this is the reason
behind the negligible loop effects for $g_X \lesssim 0.4$. However, we
would like to point out that this behavior is not to be generally
expected and emphasize the relevance of loop effects for a proper
evaluation of BR($\tau \to 3 \, \mu$) in $Z^\prime$ models for the $b
\to s$ anomalies.

\section{Summary and conclusions}
\label{sec:conclusions}

The hints reported by the LHCb collaboration may be the first
indications of a completely unexpected New Physics sector with
interactions that violate lepton flavor universality. In this paper we
have explored an extension of the $Z^\prime$ model of
\cite{Sierra:2015fma} with a non-trivial embedding of neutrino masses
and mixings. Our focus has been on the lepton flavor violating
phenomenology of the resulting model, motivated by theoretical
arguments that link it to the breaking of lepton flavor universality
\cite{Glashow:2014iga}.

The main conclusions of our phenomenological exploration can be
summarized as follows:

\begin{itemize}
\item The additional degrees of freedom introduced to accommodate
  neutrino masses and mixings play a sub-dominant role in the lepton
  flavor violating predictions of the model, which are dominated by
  the New Physics effects induced by the states responsible for the
  explanation of the $b \to s$ anomalies.
\item In most parts of the parameter space the rates for $B \to K \tau
  \mu$ and $\tau \to 3 \, \mu$ are strongly correlated. This is simply
  due to the fact that both are dominated by tree-level $Z^\prime$
  boson exchange. In this case, we have derived the upper limit
  $\text{BR}(B \to K \tau \mu)_{\text{max}} \lesssim 8 \cdot
  10^{-10}$. This limit applies to all models with purely left-handed
  $Z^\prime$ couplings and can only be evaded by fine-tuning the
  contributions to $B_s - \overline{B_s}$ mixing in models with both
  left- and right-handed $Z^\prime$
  couplings~\cite{Crivellin:2015era}.
\item Loop effects in $\tau \to 3 \, \mu$ may be comparable to the
  tree-level ones. This is due to the strong suppression induced by
  the tree-level exchange of a TeV-scale $Z^\prime$ boson, which is
  absent in many 1-loop contributions. In fact, this feature is
  expected in generic $Z^\prime$ models for the $b \to s$ anomalies,
  although some regions of the parameter space of these models might
  deviate from this general expectation.
\end{itemize}

Flavor processes are clearly the most direct test of the model under
discussion and crucial contributions from the Belle II experiment are
expected in the long term \cite{Albrecht:2017odf}. However, the model
can also be probed in several complementary ways. Direct searches at
the LHC can also provide an additional handle on the model. One can
have observable production rates for the vector-like lepton in the
model, see \cite{Xu:2018} for a recent work in this direction, or
search for the mediator of the New Physics contributions, the heavy
$Z^\prime$ boson, see for instance \cite{Faroughy:2016osc}. If the $b
\to s$ anomalies and the violation of flavor universality are finally
confirmed, all these experimental approaches will be necessary to have
a global picture of the new dynamics that lies beyond the Standard
Model.

\section*{Acknowledgements}

The authors are grateful to M. Lucente and D. Aristizabal Sierra for
fruitful discussions. Work supported by the Spanish grants
SEV-2014-0398 and FPA2017-85216-P (AEI/FEDER, UE), SEJI/2018/033
(Generalitat Valenciana) and the Spanish Red Consolider MultiDark
FPA2017‐90566‐REDC. PR acknowledges support by CONACYT becas en el
extranjero CVU 468534 and the Bonn-Cologne graduate school (BCGS).

\providecommand{\href}[2]{#2}
\end{document}